\documentclass[intlimits,twoside,a4paper]{article}

\usepackage{amsmath,amssymb}
\usepackage{graphicx}

\usepackage[T2A]{fontenc}
\usepackage[cp1251]{inputenc}

\usepackage[eqsecnum]{cmpj3}

\issue{2017}{20}{2}{23702}
\doinumber{10.5488/CMP.20.23702}
\title{The perfect spin injection in silicene FS/NS junction}
\author[H.-Y. Tian, N. Xu, G. Luo, Ch.-D. Ren]{H.-Y. Tian\refaddr{label1}, N. Xu\refaddr{label1}, G. Luo\refaddr{label1}, Ch.-D. Ren\refaddr{label2}\thanks{E-mail: renchongdan@hotmail.com}}
\addresses{
\addr{label1} Department of Physics, Yancheng Institute of Technology, Jiangsu 224051, China
\addr{label2} Department of Physics, Zunyi Normal Colleage, Guizhou 563002, China
}

\date{Received February 17, 2017, in final form April 2, 2017}

\DeclareMathOperator{\Tr}{Tr}

\begin{document}

\maketitle

\begin{abstract}
We theoretically investigate the spin injection from a ferromagnetic silicene to a normal silicene
(FS/NS), where the magnetization in the FS is assumed from the magnetic proximity
effect. Based on a silicene lattice model, we demonstrated that the pure spin injection could be
obtained by tuning the Fermi energy of two spin species, where one is in the spin orbit coupling gap
and the other one is outside the gap.
Moreover, the valley polarity of
the spin species can be controlled by
a perpendicular electric field in the FS region.
Our findings
may shed light on making silicene-based spin and valley devices in the spintronics
and valleytronics field.
\pacs 73.43.-f, 73.43.Nq, 72.80.Ey
\end{abstract}

\section{Introduction}
Silicene, monolayer of silicon with low-buckled structure,
is one of the most fascinating Dirac materials following graphene \cite{1,2,3}. Silicene possesses almost every remarkable properties with graphene, such as two inequivalent valleys $(K,K')$ at the corners of the first Brillioun zone,
a linear band structure near the $K$ $(K')$ points \cite{4,5} as well as the Klein paradox \cite{6}.
In contrast to graphene, silicene has a
large spin-orbit coupling (SOC) due to the low-buckled geometry which
opens a gap between the conduction and valence bands, and the energy gap can be further tuned
by an external electric field perpendicular to the silicene sheet \cite{7,8}.
Recently, the silicene nanoribbon has been synthesized
on Ag surface \cite{9,10}, and the silicene field effect
transistor has also been successfully fabricated in experiment \cite{11}.
Besides the quantum spin Hall effect resulting from the strong SOC where the two spin species counter-propagate along one
sample edge, the silicene has a long spin
dephasing length at room temperature \cite{12}. Furthermore, it can easily be incorporated
into the traditional silicon-based nanotechnology \cite{13},
which makes a silicene a good candidate for the prospective spintronics.

A principal challenge in making silicene based spin devices is to realize a highly efficient spin injection from
a spin resource to a silicene.
At present, the central issue of a spin transport in a silicene is mainly focused
on the manipulation of the spin polarized current,
taking into account the edge state regime \cite{2,7,14,15,16} or the bulk spin \cite{17,18,19} by introducing
a Zeeman field and a ferromagnetic gate to make the
bulk silicene ferromagnetic locally.
However, the spin injection from a spin resource to a silicene like
that in graphene \cite{20} is rarely discussed theoretically and experimentally,
although some proposals were implemented
in silicon \cite{21,22,23,24}.
As concerns theoretic aspects, the best choice to realize a highly efficient spin injection
is to construct a tunnel barrier in order to inhibit one spin species.
In our previous work, we created an artificial tunnel barrier, i.e., a lattice void,
at the interface of the ferromagnetic/normal graphene (FG/NG)
and realized a nearly pure spin injection in the NG region along
with the total internal reflection phenomenon \cite{25}.
In comparison with graphene, silicene has a relatively large
energy gap which provides a natural tunnel barrier and can prevent
all the bulk electrons, so that the unitary spin injection can be achieved in the
ferromagnetic/normal silicene (FS/NS) junction.
Moreover, the two valleys $K$ and $K'$ can be distinguishable when the SOC
and perpendicular electric field are both present which makes it possible
to control the valley character of the spin species in the NS region.

\begin{figure}[!t]
\begin{center}
\includegraphics[scale=0.35]{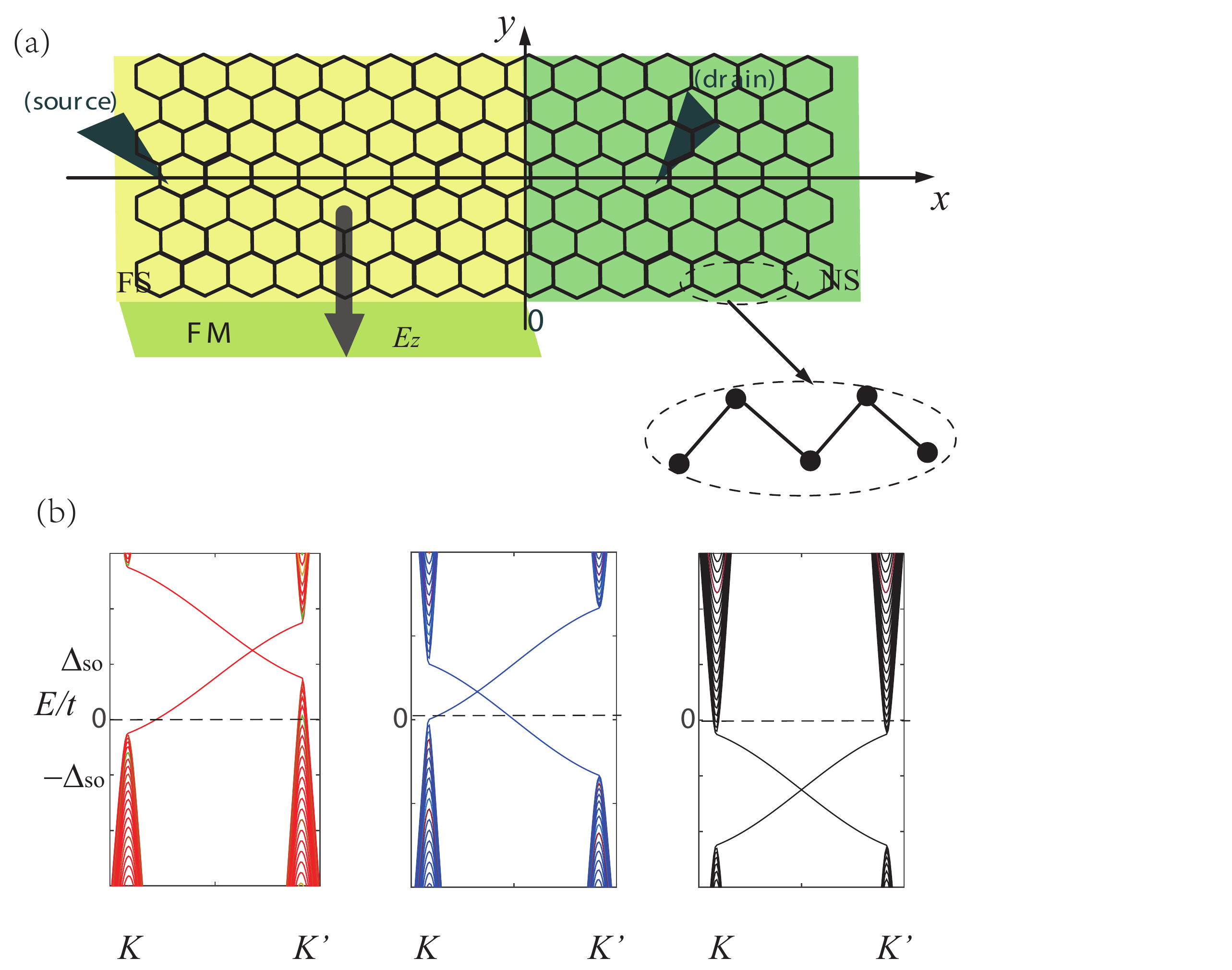}
\end{center}
\caption{\label{fig1} (Color online) Construction of the silicene FS/NS junction studied.
(a) Real-space lattice structure, where the FS region is placed
on a ferromagnetic insulator that provides a
ferromagnetic exchange field, and is manipulated by a perpendicular
electric field $E_z$.
A point source injects electrons in the FS region and an extended drain collects electrons that scattered
from the interface in the NS region.
(b) Band structures for spin up (red) and spin down (blue) electrons in the FS region and electrons (black) in the
NS region of zigzag ribbons, the dashed line represents the Fermi energy. }
\end{figure}

In this work, we theoretically study the spin injection from the
FS into NS where the electrons are injected into the FS by a
point source and are collected in the NS by a drain electrode. A
uniform ferromagnetism in the FS is assumed to stem from the
magnetic proximity effect with a magnetic insulator supporting
the silicene as shown in figure~\ref{fig1}.
Based on a tight-binding lattice model, we demonstrated that
the perfect spin injection with $100\%$ efficiency can be realized by tuning the gate
voltage in the FS. Moreover, the spin species can originate
from one definite valley when a perpendicular electric field exists.

\section{Model and formalism}

A two-terminal FS/NS device is schematically shown in
figure~\ref{fig1}, where the left-hand semi-infinite silicene nanoribbon is
deposited on an ferromagnetic insulator that can induce an exchange
field $M$ in the FS region, while the right-hand one is the NS on
a ordinary nonmagnetic substrate. Meanwhile, an electric field $E_z$
is applied perpendicularly to the silicene sheet in the FS region
between the top gate and the back gate which can induce sublattice potentials.
The source electrode injects
electrons into the FS while the drain electrode in the NS collects
electrons.

We start with the tight-binding Hamiltonian of a silicene ribbon
including the SOC \cite{26}
\begin{align}
H&=\sum_{i\alpha}(\epsilon_{j}+M\alpha+\Delta_z\mu_{i})
a^{\dagger}_{i\alpha}a_{i\alpha}-t\sum_{\langle i,j\rangle\alpha}a^{\dagger}_{i\alpha}a_{j\alpha}+\ri\frac{\lambda_{\text{SO}}}{3\sqrt{3}}\sum_{\langle\langle i,j\rangle\rangle\alpha\beta}\nu_{ij}a_{i\alpha}^
{\dagger}\sigma_{\alpha\beta}^{z}a_{j\beta} \nonumber\\
&\quad+\sum_{\zeta,k\alpha}\epsilon_{\zeta,k}d_{\zeta,k\alpha}^{\dag}d_{\zeta,k\alpha}+(\gamma c_{i_{\zeta},\alpha}^{\dag}d_{\zeta,k\alpha}+\text{c.c.}),
\label{1}
\end{align}
where $a_{i\alpha}^{\dag}$ $(a_{i\alpha})$ is the creation (annihilation) operator for an electron with spin
$\alpha$ $(\alpha=\pm=\uparrow\downarrow)$ on site $i$,
brackets $\langle\dots\rangle$ and $\langle\langle\dots\rangle\rangle$
denote the summations over all the nearest and next-nearest neighbor sites. In the first term,
$\epsilon_{j}=\epsilon_{\text{FS(NS)}}$ denotes the on site energy of the FS (NS) region
which can be controlled by varying the gate voltage or by doping the
underlying substrate,
$M$ is the exchange field in the FS region and the last one denotes
the stagger on-site potential by a perpendicular electric field
with the stagger potential amplitude $\Delta_z$,
in which $\mu_i=-1$ $(+1)$ when $i$ represents a buckled up (down) atom.
The second and third terms represent the nearest and
next-nearest coupling with the hopping energy $t$, and the effective
SOC strength $\lambda_{\text{SO}}$. $\sigma=(\sigma_x,\sigma_y,\sigma_z)$ is
the pauli matrix of spin, $\nu_{ij}=+1$ and $\nu_{ij}=-1$
correspond to the anticlockwise and clockwise
next-nearest-neighboring hopping with respect to the positive
$z$ axis, respectively.
The last term in equation~(\ref{1}) represents the Hamiltonian of
the source and the detector leads described in the $k$ space and
their coupling to the silicene lattices $i_{\zeta}$.
Here, $\zeta=\text{s,\,d}$ represent the source and the detecting electrodes and $d_{\zeta}$ $(d_{\zeta}^{\dag})$ is the
annihilation (creation) operator of the electrons in the electrode $\zeta$.

From equation~(\ref{1}) we get to know that a silicene will exhibit
interesting topological properties
and a discussion on its features is quite helpful prior to the numerical calculations.
Figure~\ref{fig1}~(b) shows the band structure in the FS
and NS region with the zigzag edge. The most notable characteristic
is the SOC gap which reads $\Delta_{\text{SO}}=\lambda_{\text{SO}}$,
between which the edge state bands connect the
electron and the hole dispersions of different valleys.
When the exchange field is involved, the two spin bands move up and down,
so that one spin can be in the gap while the other one can lie in the
conduction (valence) band by tuning the gate voltage.
If the stagger potential is present, the energy gaps for the two
valleys are not identical anymore and they are resolved, assuming $\Delta_{K/K'}=\vert\Delta_z\pm\Delta_{\text{SO}}\vert$
for spin up and the reverse for spin down.
As $\Delta_z=\Delta_{\text{SO}}$, the gap closes, but
when it increases again the gap revives and enlarges
and the system turns into a band insulator.
For armchair ribbons, the two valleys can also become distinguishable although they
overlap with each other under the perpendicular electric field.

When the flow of electrons is injected into a silicene
in the FS region from a point electrode, the response signal is
induced everywhere in the NS region.
In order to investigate the spin injection efficiency in the silicene FS/NS junction, we will
consider the response of the local particle density to the electrons injected
from the source electrode.
In a small bias limit and at zero temperature, the variation of the local particle density of spin $\alpha$ $\delta\rho_{\alpha}(l)$
per voltage is given by \cite{27}
\begin{eqnarray}
\delta\rho_{\alpha}(l)=\Tr[G^{\text r}\Gamma_{\text s}G^{\text a}]_{l\alpha l\alpha}\,,
\label{2}
\end{eqnarray}
where the trace is over the local
lattice sites coupling with the drain electrode.
$G^{\text r}$ $(G^{\text a})$ is the retarded (advanced) Green's function and is given by \cite{28}
\begin{eqnarray}
G^{\text r}=[EI-\widetilde{H}-\Sigma_{\text s}^{\text r}-\Sigma_{\text d}^{\text r}]^{-1},
\end{eqnarray}
where $E$ is the Fermi energy, $I$ is a unit matrix, $\widetilde{H}$ is the first three terms
in equation~(\ref{1}), and $\Sigma_{\text{s(d)}}^{\text r}$ is the self-energy of
the source (drain) electrodes.
$\Gamma_{\text s}=\ri(\Sigma_{\text s}^{\text r}-\Sigma_{\text s}^{\text a})$ in equation~(\ref{2}) is the linewidth function of the source
electrode. In actual calculations, we set the
average value of the six discrete sites in a unit cell of honeycomb lattice
as a real local quantity.

\section{Results and discussions}

In the numerical calculations, we set the nearest-neighbor
carbon-carbon distance $a=0.386$~nm, the second nearestneighbor distance
$b=\sqrt{3}a=0.67$~nm, the hopping energy $t=1.6$~eV and the SOC strength
$\lambda_{\text{SO}}=0.04t$ as in a real silicene sample.

\begin{figure}[!t]
\begin{center}
\includegraphics[scale=0.5]{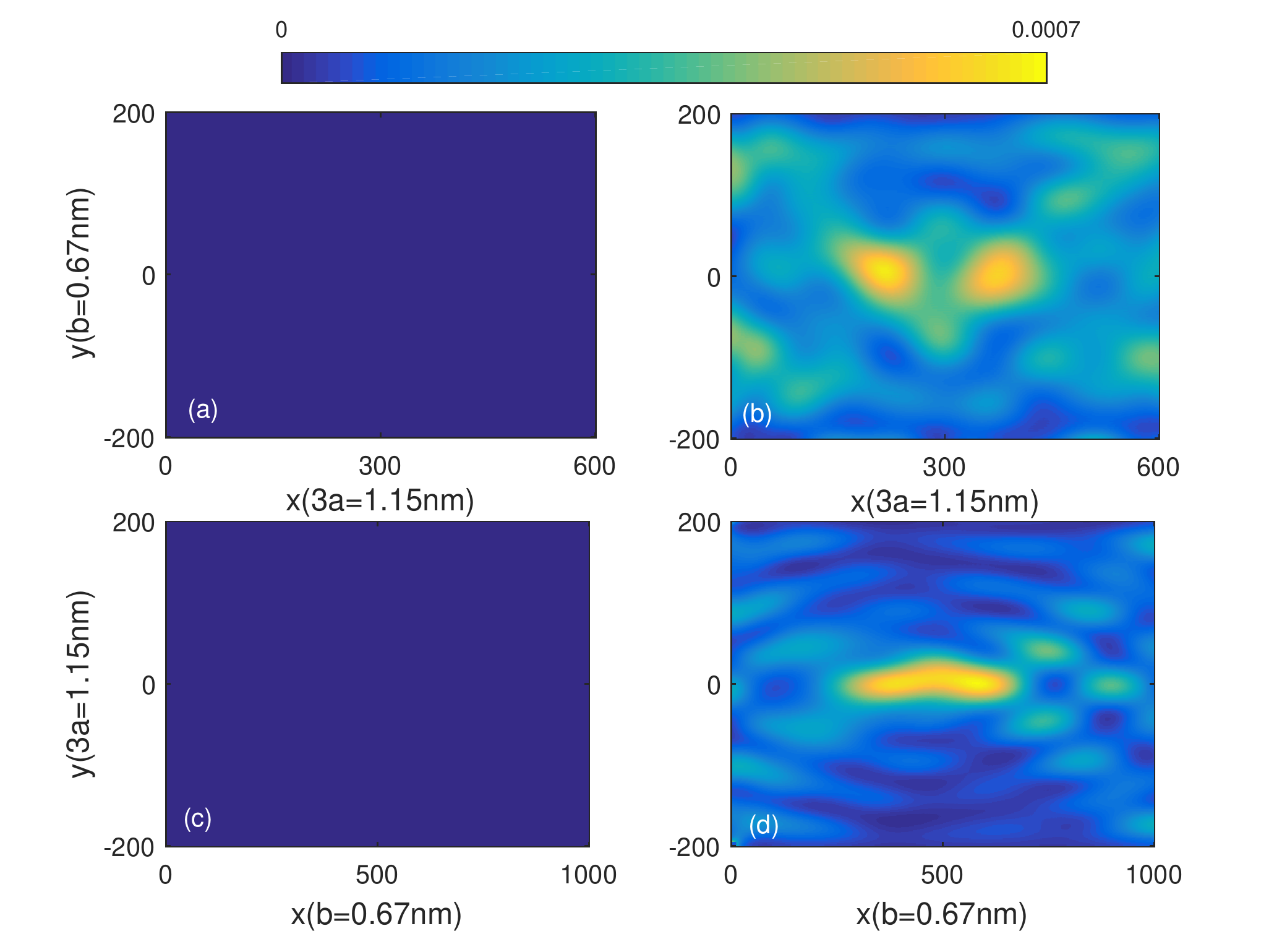}
\end{center}
\caption{\label{fig2} (Color online) Distribution of the local particle density
variation $\delta\rho_{\downarrow}(l)$ in (a) [(c)] and $\delta\rho_{\uparrow}(l)$
in (b) [(d)] with the armchair (zigzag) ribbon. The parameters are $\varepsilon_{\text{FS}}=0.03t$,
$\varepsilon_{\text{NS}}=-0.05t$, $M=0.02t$ and $\Delta_z=0$. The width of the armchair (zigzag) ribbon
is $W=401b$ ($W=401\times3a$), and the source position is at $(-300\times3a,0)$ $[(-500b,0)]$.}
\end{figure}
\begin{figure}[!b]
\vspace{-3mm}
\begin{center}
\includegraphics[scale=0.5]{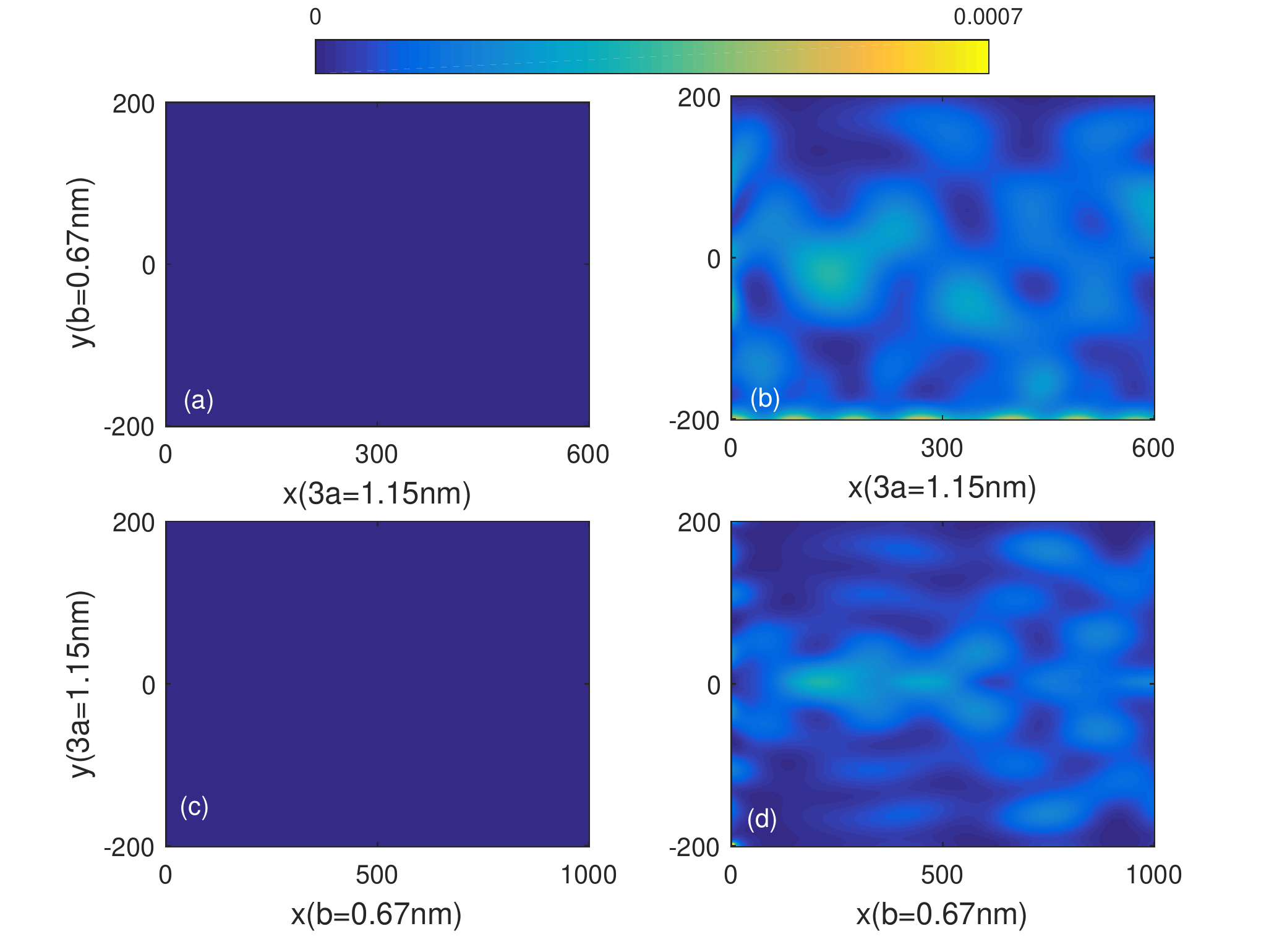}
\end{center}
\caption{\label{fig3} (Color online) Distribution of the local particle density
variation $\delta\rho_{\downarrow}(l)$ in (a) [(c)] and $\delta\rho_{\uparrow}(l)$
in (b) [(d)] with the armchair (zigzag) ribbon under a perpendicular electric field
with $\Delta_z=0.02t$. The other parameters are
the same as figure~\ref{fig2}~(a)--(d).}
\end{figure}

In figure~\ref{fig2}, we present the local particle density
variation of each spin in the NS region for an armchair and zigzag ribbon.
It is shown that for the spin down species, the $\delta\rho_{\downarrow}(l)$ approaches zero in the
whole NS region and the total quantity is of the order of $10^{-7}$.
This clearly indicates that the spin down species is nearly non-existent because
its Fermi energy lies in the energy gap. However, for the spin up species,
the $\delta\rho_{\uparrow}(l)$ has a certain quantity in the NS region and
reaches a maximum at the symmetry point of the source electrode which is caused
by the negative refraction due to the Klein tunneling effect \cite{25}.
Meanwhile, the total quantity of $\delta\rho_{\uparrow}(l)$ in the whole
NS region is 56.8 for the armchair ribbon and 63.2 for the zigzag ribbon.
Therefore, the spin up species is distributed everywhere in the NS region and the
complete spin injection is realized in the FS/NS junction.
In fact, a pure spin injection can always occur
as long as $M+\varepsilon_n>\Delta_{\text{SO}}$ and $\varepsilon_n-M\leqslant\Delta_{\text{SO}}$ so as
to ensure that the Fermi energy of the two spin species locates inside and outside the bulk gap, respectively.

In figure~\ref{fig2}~(b) and~(d), the spin up electrons scattered in the NS region originate from two different valleys
because they are still degenerate in the momentum space, meaning that one valley contributes half of the total quantity
of $\delta\rho_{\uparrow}(l)$ in the NS region.
As mentioned above, the presence of an electric field
can adjust the bulk gap width of each valley. For instance, the bulk gap at the $K$ $(K')$ valley
of spin up (down) enlarges but it shrinks at $K'$ $(K)$ valley, as shown in figure~\ref{fig1}~(b).
When the Fermi energy locates in the gap of the $K$ valley of both spins by tuning the electric field, the spin up electrons in the
$n$ region  merely originate from the $K'$ valley, the results being shown in figure~\ref{fig3}~(b) and~(d).
As is shown, $\delta\rho_{\uparrow}(l)$ decreases as compared with figure~\ref{fig2}~(b) and~(d) in figure~\ref{fig3}~(b) and~(d), and the
total quantity of $\delta\rho_{\uparrow}(l)$ in the whole
NS region is 27.4 for the armchair ribbon and 30.8 for the zigzag ribbon
which is about half the total quantity of figure~\ref{fig2}~(b) and~(d) due to the single valley contributions.
Generally speaking, the single valley spin injection arises
when $M+\varepsilon_n-\Delta_z<\Delta_{\text{SO}}$ and $\varepsilon_n-M+\Delta_z<\Delta_{\text{SO}}$,
or $M+\varepsilon_n-\Delta_z>0$ and $\vert\varepsilon_n-M-(\Delta_z-\Delta_{\text{SO}})\vert<\Delta_{\text{SO}}$
for the $\Delta_z>\Delta_{\text{SO}}$ case.

In the above, we only consider a spin injection in a bipolar silicene ribbon,
for instance, the carriers are hole-like and electron-like on both sides of the
interface. However, the results are not limited to the bipolar system
provided that one spin species lies in the bulk gap while the other one is not.
Moreover, the shape of the interface, a sharp or a smooth one,
will not affect the spin injection which makes it feasible
in a real experimental fabrication.

\section{Conclusion}
To conclude, we have studied a pure spin injection in the FS/NS junction
and found that only one spin species can emerge in the NS region and the other one
is eliminated by tuning the gate voltage in the FS region.
Moreover, the spin species can originate from one valley at a certain
electric field. The results are particularly interesting in view of
the application for spintronics,
which might be observable in doped silicene.
Our findings may provide a valuable guidance for designing and fabricating
spin devices based on a silicene ribbon in practical applications.

\section*{Acknowledgements}
This work is supported by the National Natural Science Foundation of China (Nos.~11447218, 11547189, 11447216, 11404278), the Science
Foundation of Guizhou Science and Technology Department under grant No.~QKHJZ[2015]2150, and the Science Foundation of Guizhou Provincial Eduction Department under grant No.~QJHKYZ[2016]092.

\ukrainianpart
\title{Досконала спінова інжекція на переході FS/NS силіцен }
\author{Г.-Я. Тьян\refaddr{label1}, Т. Хю\refaddr{label1}, Г. Люо\refaddr{label1}, Ч.-Д. Рен\refaddr{label2}}
\addresses{
\addr{label1} Фізичний факультет, Технологічний інститут Янченг, Цзянсу 224051, Китай
\addr{label2}  Фізичний факультет, Нормальний коледж Зунию, Гуйжоу 563002, Китай
}

\makeukrtitle

\begin{abstract}
Теоретично вивчається спінова інжекція з феромагнiтного силiцену в нормальний
 силiцен (FS/NS перехід), коли намагнiченiсть в FS припускається з магнiтного
 ефекту близькостi. На основi граткової моделi силiцену показано, що чисто
 спінова інжекція може бути отримана підлаштуванням енергiй Фермi спінів двох
 сортів, коли один сорт є в зонi спiн-орбiтальної взаємодiї, а iнший поза
 зоною. Крiм того, долинова полярнiсть спiнових сортiв може контролюватися
 перпендикулярно напрямленим електричним полем в FS областi. Нашi результати
 можуть пролити свiтло на створення на основі силiцену спiнових i долинових
 пристроїв для спiноелектронiки i велiтронiки.
\keywords силіцен, спінова інжекція, FS/NS переходи
\end{abstract}


\begin{thebibliography}{99}
\bibitem{1} Vogt P., De Padova P., Quaresima C., Avila J., Frantzeskakis E., Asensio M.C., Resta A., Ealet B., Le Lay G., Phys. Rev. Lett., 2012, \textbf{108}, No.~15, 155501, \doi{10.1103/PhysRevLett.108.155501}.
\bibitem{2} Ezawa M., Phys. Rev. Lett., 2013, \textbf{110}, No.~2, 026603, \doi{10.1103/PhysRevLett.110.026603}.
\bibitem{3} Wang J., Deng S., Liu Z., Liu Z., Natl. Sci. Rev., 2015, \textbf{2}, No.~1, 22, \doi{10.1093/nsr/nwu080}.
\bibitem{4} Cahangirov S., Topsakal M., Akt\"urk E., \c{S}ahin H., Ciraci S., Phys. Rev. Lett., 2009, \textbf{102}, No.~23, 236804, \doi{10.1103/PhysRevLett.102.236804}.
\bibitem{5} Xu Y., Yan B., Zhang H.-J., Wang J., Xu G., Tang P., Duan W., Zhang S.-C., Phys. Rev. Lett., 2013, \textbf{111}, No.~13,
136804, \doi{10.1103/PhysRevLett.111.136804}.
\bibitem{6} Hajati Y., Rashidian Z., AIP Adv., 2016, \textbf{6}, No.~2, 025307, \doi{10.1063/1.4942043}.
\bibitem{7} Ezawa M., Phys. Rev. Lett., 2012, \textbf{109}, No.~5, 055502, \doi{10.1103/PhysRevLett.109.055502}.
\bibitem{8} Ezawa M., New J. Phys., 2012, \textbf{14}, No.~3, 033003, \doi{10.1088/1367-2630/14/3/033003}.
\bibitem{9} Aufray B., Kara A., Vizzini S., Oughaddou H., L\'eandri C., Ealet B., Le Lay G., Appl. Phys. Lett., 2010, \textbf{96},
No.~18, 183102, \doi{10.1063/1.3419932}.
\bibitem{10} De Padova P., Quaresima C., Ottaviani C., Sheverdyaeva P.M., Moras P., Carbone C., Topwal D., Olivieri~B., Kara~A., Oughaddou H.,
 Aufray B., Le~Lay G., Appl. Phys. Lett., 2010, \textbf{96}, No.~26, 261905, \doi{10.1063/1.3459143}.
\bibitem{11} Tao L., Cinquanta E., Chiappe D., Grazianetti C., Fanciulli M., Dubey M., Molle A., Akinwande D., Nat. Nanotechnol.,
2015, \textbf{10}, No.~3, 227, \doi{10.1038/nnano.2014.325}.
\bibitem{12} Bishnoi B., Ghosh B., RSC Adv., 2013, \textbf{3}, No.~48, 26153, \doi{10.1039/c3ra43491e}.
\bibitem{13} Yamakage A., Ezawa M., Tanaka Y., Nagaosa N., Phys. Rev. B, 2013, \textbf{88}, No.~8, 085322,\\ \doi{10.1103/PhysRevB.88.085322}.
\bibitem{14} Liu C.-C., Feng W., Yao Y., Phys. Rev. Lett., 2011, \textbf{107}, No.~7, 076802, \doi{10.1103/PhysRevLett.107.076802}.
\bibitem{15} Pan H., Li Z., Liu C.-C., Zhu G., Qiao Z., Yao Y., Phys. Rev. Lett., 2014, \textbf{112}, No.~10, 106802, \doi{10.1103/PhysRevLett.112.106802}.
\bibitem{16} Ezawa M., Phys. Rev. B, 2013, \textbf{87}, No.~15, 155415, \doi{10.1103/PhysRevB.87.155415}.
\bibitem{17} Tsai W.-F., Huang C.-Y., Chang T.-R., Lin H., Jeng H.-T., Bansil A., Nat. Commun., 2013, \textbf{4}, 1500, \doi{10.1038/ncomms2525}.
\bibitem{18} Gupta G., Lin H., Bansil A., Jalil M.B.A., Huang C.-Y., Tsai W.-F., Liang G., Appl. Phys. Lett., 2014, \textbf{104},
No.~3, 032410, \doi{10.1063/1.4863088}.
\bibitem{19} Yokoyama T., Phys. Rev. B, 2013, \textbf{87}, No.~24, 241409, \doi{10.1103/PhysRevB.87.241409}.
\bibitem{20} Han W., Wang W.H., Pi K., McCreary K.M., Bao W., Li Y., Miao F., Lau C.N., Kawakami R.K., Phys. Rev. Lett.,
2009, \textbf{102}, No.~13, 137205, \doi{10.1103/PhysRevLett.102.137205}.
\bibitem{21} Ando Y., Hamaya K., Kasahara K., Kishi Y., Ueda K., Sawano K., Sadoh T., Miyao M., Appl. Phys. Lett., 2009,
\textbf{94}, No.~18, 182105, \doi{10.1063/1.3130211}.
\bibitem{22} Li C.H., van 't Erve O.M.J., Jonker B.T., Nat. Commun., 2011, \textbf{2}, 245, \doi{10.1038/ncomms1256}.
\bibitem{23} Tang J., Wang K.L., Nanoscale, 2015, \textbf{7}, No.~10, 4325, \doi{10.1039/C4NR07611G}.
\bibitem{24} Zhang S., Dayeh S.A., Li Y., Crooker S.A., Smith D.L., Picraux S.T., Nano Lett., 2013, \textbf{13}, No.~2, 430, \doi{10.1021/nl303667v}.
\bibitem{25} Tian H.Y., Chan K.S., Wang J., Phys. Rev. B, 2012, \textbf{86}, No.~24, 245413, \doi{10.1103/PhysRevB.86.245413}.
\bibitem{26} Yang M., Song X.L., Chen D.H., Bai Y.K., Phys. Lett. A, 2015, \textbf{379}, No.~16--17, 1149,\\ \doi{10.1016/j.physleta.2015.02.021}.
\bibitem{27} Tao W.W., Liu B., Dai Q., Wang S.-K., Commun. Theor. Phys., 2014, \textbf{61}, No.~3, 391,\\ \doi{10.1088/0253-6102/61/3/20}.
\bibitem{28} Xing Y., Wang J., Sun Q.-F., Phys. Rev. B, 2010, \textbf{81}, No.~16, 165425, \doi{10.1103/PhysRevB.81.165425}.
\end{thebibliography}
 \end{document}